\documentclass[12pt]{article} 
\usepackage{amsmath}
\usepackage{amsfonts,amssymb}
\usepackage[]{hyperref}

\newtheorem{thm}{Theorem}[section]

\newtheorem{rem}{Remarks}[section]
\newtheorem{claim}{Claim}[section]
\numberwithin{equation}{section}

\begin{document}

\title { On the recent paper on quark confinement by Tomboulis 
       }
\author { K. R. Ito
          \thanks{E-mail : ito@mpg.setsunan.ac.jp} \\
           Department of Mathematics and Physics\\
           Setsunan University\\
          Neyagawa, Osaka 572-8508, Japan\\
          E. Seiler
          \thanks{E-mail : ehs@mppmu.mpg.de} \\
          Max-Planck Institut f\"ur Physik\\
          F\"ohringer Ring 6\\
          D-80805 M\"unchen, Germany\\
          }
\date {November 30, 2007}

\maketitle

\begin{abstract}
      We point out missing links in the recent paper 
      by Tomboulis in which he claims a rigorous proof of quark 
      confinement in 4D lattice gauge  theory. 
      We also discuss if it is possible to correct his proof.\\

     \hspace*{2cm}  quark confinement,  Migdal-Kadanoff type 
\end{abstract}

\section {Introduction}

Many physicists have been fascinated by the problem of quark confinement 
\cite{wil} that remains as one of the most mysterious problems in modern 
physics since the last century. It is still an open problem if lattice 
gauge theory based on a non-abelian gauge group confines quarks for all 
values of the (bare) coupling constant, and if a continuum limit exists in 
which quark confinement and asymptotic freedom coexist (this has been 
questioned by A. Patrascioiu and one of the present authors, see 
\cite{seiler} and references therein).

Recently E.~T.~Tomboulis published a paper in the arXiv \cite{tomb} in 
which he claims to present a rigorous proof of quark confinement in 4D 
lattice gauge theory. We think that though the idea in \cite{tomb} may be 
interesting, the proof in \cite{tomb} depends on an assumption or a claim 
which remains to be proved. We also discuss if it is possible to correct 
his proof.

\section {Tomboulis's alleged theorem}

We summarize his notation and arguments with some simplifications:

\begin{enumerate}
\item  $\Lambda \subset Z^{4}$ is the periodic box in $Z^{4}$ of size 
       $L_{1}\times \cdots \times L_{4}$ with center at the origin, 
       where $L_{i}=ab^{n_{i}}$, $n_{i}>>1$, $i=1,\cdots,4$ and 
       $a,b$ are positive integers larger than 1.
\item  $\Lambda^{(k)} \subset Z^{4}$ is the periodic box in $Z^{4}$ 
       of each side length $L_{i} b^{-k}$ obtained from $\Lambda$ 
       by $k$ steps of the renormalization transformation of 
       Migdal-Kadanoff type ($\Lambda=\Lambda^{(0)}$). 
\item  $f(\{c_{j}\},  U)=1+\sum_{j\neq 0} c_{j}d_{j}\chi_{j}(U)$ 
        where $d_{j}$ is the dimension of the representation $\chi_{j}$, 
        and we assume $U\in G=SU(2)$. Moreover 
        $U=U_{p}=\prod_{b\in\partial p} U_{b}$ for a plaquette 
        $p\subset \Lambda$ which consists of four oriented bonds  
        $b\in \partial p$
\end{enumerate}

We start with
\begin{subequations}
\begin{eqnarray}
\exp\left [ \frac{\beta}{2}\chi_{1/2}(U) \right ]
       &=&F_{0}(0)f(\{c_{j}\}, U) \\
f(\{c_{j}\},U) &=& 1+\sum_{j\neq 0} c_{j}d_{j}\chi_{j}(U) \\
F_{0}(0) &=& \int \exp\left[  \frac{\beta}{2} \chi_{1/2}(U)\right ] dU
\end{eqnarray}
\end{subequations}
where $dU$ is the Haar measure of $G=SU(2)$. We then apply the
renormalization group (RG) recursion formulas of Migdal-Kadanoff type 
 \cite{mk} with correction parameters. If there are no correction 
parameters (the standard recursion formula), we have
\begin{eqnarray*}
f^{(n-1)}(U) &=&f(\{c_{j}(n-1)\},U) \\
             &=& 1+\sum_{j\neq 0} c_{j}(n-1)d_{j}\chi_{j}(U)\\
             &\to& f^{(n)}(U)=f(\{c_{j}(n)\},U)
\end{eqnarray*}
where
\begin{eqnarray*}
f^{(n)} (U) &=& 
            \frac{1}{F_{0}(n)} \int
             \left(f^{(n-1)}(UU_{1})
                   f^{(n-1)}(U_{1}^{-1}U_{2})
                   \cdots 
                   f^{(n-1)}(U_{b^{2}})\right)^{b^{2}}
                   \prod dU_{k} \\
    F_{0}(n) &=& \left(\int [f^{(n-1)}(U)]^{b^{2}} dU\right )^{b^{2}}
\end{eqnarray*}
or equivalently 
\begin{eqnarray}
c_{j}(n) &=& 
            \frac{F_{j}(n)}{F_{0}(n)} \\
    F_{j}(n) &=&
           \left( \int [f^{(n-1)}(U)]^{b^{2}}\frac{\chi_{j}(U)}{d_{j}}
            dU \right)^{b^{2}}
\end{eqnarray}
in terms of the coefficients of the character expansions. Then  
\cite{ito}: 
\begin{thm} For $D\leq 4$ and for $G=SU(N)$ or $G=U(N)$, 
     $$\lim_{n\to\infty} c_{j}(n)=0\quad\quad {\rm for}\; j\neq 0$$
\end{thm}

These recursions are just approximative and yield upper bounds 
for the partition functions \cite{tomb}. Then two parameters
$\alpha\in(0,1]$ and  $t>0$  and a function $h(\alpha,t) \in (0,1]$ 
are introduced in \cite{tomb} so that this transformation is numerically 
exact:
\begin{subequations}
\begin{eqnarray}
Z &=& \int dU_{\Lambda} \prod_{p\subset \Lambda } f(\{c_{j}\},U_{p})  \\
  &=& [F_{0}(1)^{h(\alpha,t)}]^{|\Lambda^{(1)}|}
      \int dU_{\Lambda^{(1)}} \prod_{p\subset \Lambda^{(1)}} 
      f(\{\tilde{c}_{j}(\alpha)\},U_{p}) \\
  &=& \tilde{Z}_{1} (\tilde{c}(\alpha),t) 
\end{eqnarray}
\end{subequations}
where $dU_{\Lambda}=\prod_{b\in\Lambda} dU_{b}$, 
\begin{subequations}
\begin{eqnarray}
\tilde{Z}_{1} (\tilde{c}(\alpha),t) 
      &=& [F_{0}(1)^{h(\alpha,t)}]^{|\Lambda^{(1)}|} Z_{1}\\
Z_{1} &=& \int dU_{\Lambda^{(1)}} \prod_{p\subset \Lambda^{(1)}} 
           f(\{\tilde{c}_{j}(\alpha)\},U_{p}) 
\end{eqnarray}
\end{subequations}
and as usual, $U_{p}=\prod_{b\in\partial p} U_{b}$
are plaquette actions defined as the product of 
group elements $U_{b}=U_{(x,x+e_{\mu})}\in G$
attached to the (oriented) bonds $b\in \Lambda$. Moreover
\begin{subequations}
\begin{eqnarray}
h(\alpha,t) &=& \exp[-t(1-\alpha)/\alpha] \\
\tilde{c}_{j}(\alpha)
            &=& \tilde{c}_{j}^{(1)}(\alpha)= \alpha c_{j}(1) \\
 c_{j}(1)   &=& \frac{F_{j}(1)}{F_{0}(1)}
\end{eqnarray}
\end{subequations}
and $\alpha=\alpha(t) \in [0,1]$ is chosen so that the above 
relation becomes exact.

Tomboulis also introduces a vortex $V=\{v\subset \Lambda\}$  \cite{mack} 
which is a collection of plaquettes 
$\{v=\{x_{0}+ne_{3},x_{0}+e_{1}+ne_{3}, x_{0}+e_{2}+ne_{3},
x_{0}+e_{1}+e_{2}+ne_{3}\}, n=0,1,,\cdots, L_{3} \}$, 
namely a plaquette $p=(x_{0}, x_{0}+e_{1}, x_{0}+e_{1}+e_{2}, 
x_{0}+e_{2})$ in an $x_{1}-x_{2}$ plane and its translations 
along one of the axis normal to $p$ (say, 3rd or 4th axis). 
We define
\begin{eqnarray}
Z^{-} &=& \int dU_{\Lambda} \prod_{p\subset \Lambda } 
           f(\{c_{j}\},(-1)^{\nu(p)}U_{p}) 
\end{eqnarray}
where 
\begin{equation}
\nu(p)=\left\{ \begin{array}{ll}
                  0  & \mbox {if $p\notin V$}\\
                  1  &  \mbox {if $p\in V$}\\
          \end{array}
         \right .
\end{equation}
and then 
\begin{eqnarray}
Z^{-} &=& \int dU_{\Lambda} \prod_{p\subset \Lambda\backslash V} 
              (1+\sum_{j\neq 0} c_{j}d_{j}\chi_{j}(U_{p}))
                     \nonumber \\
      & & \times      \prod_{q\subset V} 
              (1+\sum_{j\neq 0} (-1)^{2j}c_{j}d_{j}\chi_{j}(U_{q}))
\end{eqnarray}

Note that the position of the vortex $V\subset \Lambda$ can 
be freely moved in the $x_{1}-x_{2}$ plane by gauge 
invariance. It is easy to see that 
\begin{thm} Assume $c_{j}\geq 0$. Then \\
(1) The measure $d\mu_{\Lambda}$ defined by
$$
d\mu_{\Lambda}=\prod_{p} (1+\sum_{j\neq 0} c_{j}d_{j}\chi_{j}(U_{p}))
             dU_{\Lambda}
$$
is reflection positive with respect to all planes dividing 
the periodic box $\Lambda$. \\
(2) The measure $d\mu^{(+)}_{\Lambda}$ defined by
$$
d\mu_{\Lambda}^{+}=\left[\prod_{p} (1+\sum_{j\neq 0} c_{j}d_{j}\chi_{j}(U_{p}))
                +\prod_{p} (1+\sum_{j\neq 0} c_{j}d_{j}
                        \chi_{j}((-1)^{\nu(p)}U_{p}))
              \right ]
               dU_{\Lambda}
$$
is reflection positive with respect to all planes
without bisecting $V\subset \Lambda$
\end{thm}

\begin{rem} (1) Though the reflection positivity of interactions plays a
very important role in general, only $\{c_{j}(n) \geq 0\}$ is
used in the paper \cite{tomb}.  \\
(2) The vortex structure in $Z^{-}$ is kept in $Z^{-}_{\Lambda ^{(n)}}$ 
by the RG transformations of Migdal-Kadanoff type.
\end{rem}

The main claim in \cite {tomb} is: 
\begin{claim} There exist $t\geq 0$ such that 

$$
1+\frac{Z^{-}_{\Lambda}(\{c_{j}\})} {Z_{\Lambda}(\{c_{j}\})}
=
1+\frac{Z^{-}_{\Lambda^{(1)}}(\{\tilde{c}^{(1)}_{j}(\alpha(t))\})}
       {Z_{\Lambda^{(1)}}(\{\tilde{c}^{(1)}_{j}(\alpha(t))\})}
$$
where
$$
\tilde{c}^{(1)}_{j}(\alpha)=\alpha c_{j}(1). 
$$
\end{claim}

If this were correct, we could have 
$$
\frac{Z^{-}_{\Lambda}(\{c_{j}\})} {Z_{\Lambda}(\{c_{j}\})}
=
\frac{Z^{-}_{\Lambda^{(n)}}(\{\tilde{c}^{(n)}_{j}(\alpha(t))\})}
       {Z_{\Lambda^{(n)}}(\{\tilde{c}^{(n)}_{j}(\alpha(t))\})}
$$
by induction. Since $\{c_{j}^{(n)} \geq 0 \}$ tends to the high 
temperature fixed point(i.e. $\{c_{j}(n)\} \to 0$)  as $n\to\infty$
if the dimension is $\leq4$, whether $G$ is abelian or non-abelian
(see the remark below and \cite{ito}), this would mean strict 
positivity of 't\,Hooft's string tension and then 
establish permanent confinement of quarks in the sense of Wilson
at least for all values of the bare coupling constant \cite{mack,bose} in 
4 dimensional lattice gauge theory, thereby solving a longstanding problem 
in modern physics.

This cannot be correct, however, since there exists a deconfining 
Kosterlitz-Thouless (KT) type transition in 4D lattice gauge theory based 
on abelian gauge groups. But it is meaningful to ask why this wrong 
conclusion is reached.

\begin{rem} (1) The introduction of $0<\alpha\leq 1$ 
into $1+\sum_{j\neq 0} c_{j}d_{j}\chi_{j}(U)$ does not 
violate conditions (positivity, analyticity, class functions  etc.) 
on $f^{(n)}(v)$ in \cite{ito} since 
\begin{eqnarray*}
1+\sum_{j\neq 0} \alpha c_{j}d_{j}\chi_{j}(U)
  &=&(1-\alpha)+\alpha\left(1+\sum_{j\neq 0} c_{j}d_{j}\chi_{j}(U)\right) \\
  &=& (1-\alpha)+\alpha \left(\mbox{solution by the Migdal-Kadanoff 
formula}\right)
\end{eqnarray*}
and $0<\alpha\leq 1$. Then $\{c_{j}(n)\geq 0\}$ tends to 0 as 
$n\to\infty$. \\
(2) Tomboulis introduces another interpolation parameter $r\in(0,1]$
into the Migdal-Kadanoff formula ($b^{2}$ convolutions are replaced  
by $b^{2}r$ ). The parameter $r$ increases  the dimension D 
from $D=4$ to $D \geq 4$ from the point of view of the renormalization 
groups. So we set $r=1$ in this paper. The introduction of $r$ 
does not change our argument.\\
(3) The conjecture raised in \cite{mack} is proved rigorously in 
\cite{bose}. Namely 't\,Hooft's string tension is smaller than or equal 
to Wilson's string tension.
\end{rem}

\section {Tomboulis's  arguments revisited}

We follow arguments in \cite{tomb}. First, using the fact that 
the partition function $Z=Z_{\Lambda}$ increases by the Migdal-Kadanoff 
recursion formula \cite{tomb}, he introduces two interpolation 
parameters $\alpha$ ( $Z$ increases as $\alpha\nearrow 1$) and $t$ 
(the factor $[F_{0}(n)]^{h(\alpha,t)}$ decreases as $t$ increases). 
Then he claims that there exist functions $\alpha(t)$ and 
$\alpha^{+}(t)$ such that 
\begin{subequations}
\begin{eqnarray}
Z_{\Lambda}(\{c_{j}\})
     &=&[F_{0}(1)]^{h(\alpha(t),t)|\Lambda^{(1)}|}
     Z_{\Lambda^{(1)}}(\{\tilde{c}_{j}(\alpha(t))\}) \\
Z_{\Lambda}^{+}(\{c_{j}\})
    &=&[F_{0}(1)]^{h(\alpha^{+}(t),t)|\Lambda^{(1)}|}
     Z_{\Lambda^{(1)}}^{+}(\{\tilde{c}_{j}(\alpha^{+}(t))\})
\end{eqnarray}
\end{subequations}
where
$$ Z^{+}=\frac{1}{2}(Z+Z^{-}) $$
and the right hand sides are independent of $t$. 

The author of \cite{tomb} then claims is that there exists a $t_{*}>0$
such that 
\begin{equation}
 \alpha (t_{*})=\alpha^{+} (t_{*})  
                   \label{XXX}
\end{equation}
which yields
\begin{equation}
\frac{Z_{\Lambda}^{+} (\{c_{j}\})} {Z_{\Lambda}(\{c_{j}\})}
=\frac{Z_{\Lambda^{(1)}}^{+} (\{\tilde{c}_{j}\})}
      {Z_{\Lambda^{(1)}} (\{\tilde{c}_{j}\})}
\end{equation}
where
$$  \tilde{c}_{j}= \tilde{c}_{j}(\alpha(t_{*}))$$

But this is not proved in the paper \cite{tomb}. 
We follow his arguments by introducing the derivatives of the 
free energies with respect to $\alpha$: 
\begin{subequations}
\begin{eqnarray}
A(\alpha) &=& \frac{1}{\log(F_{0}(1))|\Lambda^{(1)}|}
      \frac{\partial}{\partial \alpha}
       \log  Z_{1}(\{\tilde{c}_{j}(\alpha)\})\\
A^{+}(\alpha) &=& \frac{1}{\log(F_{0}(1))|\Lambda^{(1)}|}
      \frac{\partial}{\partial \alpha}
       \log Z_{1}^{+}(\{\tilde{c}_{j}(\alpha)\})
\end{eqnarray}
\end{subequations}
(Here and hereafter we write $Z_{n}=Z_{\Lambda^{(n)}}$.)
Then the author of \cite{tomb} proves that if 
\begin{equation}
A(\alpha)> A^{+}(\alpha) \label{YYY}
\end{equation}
then equation (\ref{XXX}) has the solution. Namely equation
(\ref{XXX}) is reduced to inequality (\ref{YYY}) which is, 
as far as we can see, not proven, even though the author remarks 
at the beginning of page 27 of \cite{tomb}: 

\begin{quotation}
Assume now that under successive decimations the coefficients 
$c_{j}^{U}(m)$ evolve within the convergence radius $\cdots$.
Taking then $n$ sufficiently large, we need establish inequality
(5.15) (namely $A\geq A^{+}$) only at strong coupling. 
\end{quotation}
\noindent
Furthermore the discussion soon after ineq.(5.6) in \cite{tomb}
is written as if equation (\ref{XXX}) were trivial or proven,
and the following equation ((5.21) in \cite{tomb}) is written
without any explanation:
\begin{quotation}
  $$ \cdots =\frac{A^{(-)}(\xi)}{A(\xi)} \leq 1 
     \mbox{\hspace*{3cm} (5.21)}$$ 
\end{quotation}
{\it It is not clear where and how his claim is proven for large $\beta$ 
where the high-temperature expansion never works!}

Inequality (\ref{YYY}) is not trivial at all since it involves
derivatives of $\log$'s of presumably large functions. 
(This inequality is obvious when $\{c_{j}\geq 0\}$ are small and
the formal expansion converges. )

The implicit function theorem is used in \cite {tomb} to prove the 
existence of $t=t_{*}$ satisfying (\ref{XXX}).
As in \cite{tomb}, we introduce

\begin{eqnarray*}
&& \Psi(\lambda,t)= h(\alpha(t),t)
                   +\frac{1}{\log(F_{0}(1))|\Lambda^{(1)}|}
                   \nonumber \\
&& \times \left((1-\lambda)
                     \log Z_{1}^{+}(\{\tilde{c}_{j}(\alpha^{+}(t))\})
                   +\lambda \log Z_{1}^{+}(\{\tilde{c}_{j}(\alpha(t))\})
                   -\log Z^{+}(\{c_{j}\}) \right)
\end{eqnarray*}
where
\begin{eqnarray*}
\log Z^{+}(\{c_{j}\})
   &=& \log\left[F_{0}(1)^{h(\alpha^{+}(t),t)|\Lambda^{(1)}|}
                 Z_{1}^{+}(\{\tilde{c}(\alpha^{+}(t))\})\right]\\
   &=& \log\left[F_{0}(1)^{h(\alpha^{+}(t_{I}),t)|\Lambda^{(1)}|}
                 Z_{1}^{+}(\{\tilde{c}(\alpha^{+}(t_{I}))\})\right]
\end{eqnarray*}
by the parametrization invariance ($t$-invariance) of the 
partition function (this is the definition of $\alpha^{+}$).
Then
\begin{subequations}
\begin{eqnarray}
&& \Psi(\lambda=0,t) = h(\alpha(t),t)-h(\alpha^{+}(t_{I}),t_{I})\\
&& \Psi(\lambda=1,t)= 
         \frac{1}{\log F_{0}(1) |\Lambda^{(1)}|} \nonumber \\
&&\times 
         \left(\log \left[F_{0}(1)^{h(\alpha(t),t)|\Lambda^{(1)}|}
                 Z_{1}^{+}(\{\tilde{c}_{j}(\alpha(t))\})\right ]
                 -\log Z^{+}(\{c_{j}\}) \right )
\end{eqnarray}
\end{subequations}
We can assume that the equation $\Psi(\lambda=0, t)=0$ is solved 
by $t=t_{0}$, and the equation $\Psi(\lambda,t)=0$ is our required 
equation, and we want to know if the solution $t=t(\lambda)$ with 
$t(0)=t_{0}$ can be continued to $t(1)$.  We have
\begin{eqnarray}
t(\lambda)&=& t_{0}+ \int_{0}^{\lambda}
                F(s,t(s)) ds \\
F(s,t(s))   &=& - \frac{\Psi_{s}(s,t(s))}{\Psi_{t}(s,t(s))}
\end{eqnarray}
which can be analytically solved by iteration if 
$F(s,t)$ is bounded in the region. Here
\begin{eqnarray}
\Psi_{t}(\lambda,t) &=&\left[1-
     \frac{h_{\alpha}(\alpha,t)+\lambda A^{+}(\alpha)}
          {h_{\alpha}(\alpha,t)+A(\alpha)} \right]
                     h_{t}(\alpha,t) \\
\Psi_{\lambda}(\lambda,t)
       & = &   \frac{1}{\log F_{0}(1) |\Lambda^{(1)}|} \nonumber \\
       & & \times
           \left(\log  Z_{1}^{+}(\{\tilde{c}_{j}(\alpha(t))\})
             -\log Z^{+}(\{\tilde{c}_{j}(\alpha^{+}(t)\}) \right )
\end{eqnarray}
and 
\begin{equation}
   h_{t}(\alpha,t)=-\frac{1-\alpha}{\alpha} h(\alpha,t)
\end{equation}
So if $A^{+}(\alpha) > A(\alpha)$  ($0<\alpha <  1$), then 
$\Psi_{t}(\lambda,t)=0$ for some $0<\lambda < 1$ and 
the integrand $F(s,t)$ diverges at  some $s=s_{0}<1$. 
Thus we cannot expect that the solution can be continued to yield $t(1)$.

As is pointed out in \cite{tomb} and as is easily proved, 
we can prove $A >  A^{+}$ if 
$\{c_{j}\geq 0\}$ are small and the high-temperature 
expansion converges. But we do not see that the proof of his 
claim for large $\beta$ is given in the paper \cite{tomb}.

\section {Discussion}

If the conventional wisdom of quark confinement in 4D non-abelian 
lattice gauge theory is correct, 
the alleged theorem in \cite{tomb} may hold for $G=SU(N)$.
But it is again a very subtle problem to show the existence of $t$ 
such that $\alpha(t)=\alpha^{+}(t)$ since it does not exist 
in the case of $G=U(1)$.

Though the Migdal-Kadanoff RG recursion formulas cannot distinguish 
non-abelian groups from  abelian ones, the velocities of the 
convergences of $\{c_{j}(n)\}_{j=1/2}^{\infty}$ to 0  as $n\to\infty$ 
are very different. We are rather skeptical about the idea that the 
problem of quark confinement can be solved by soft analysis like this, 
but {\it if} the Migdal-Kadanoff RG formulas should play a role in a 
rigorous proof of quark confinement in lattice gauge theory, this
fact would certainly have to come into play.

In the case of $D=3$, in which case $\{c_{j}(n)\}_{j=1/2}^{\infty}$ 
converges to $0$ exponentially fast as $n\to\infty$, we may have a 
chance to apply his idea to the problem of quark confinement in 3D 
lattice gauge theory which is not yet solved. But so far, we do not
know the method. \\

{\it {\bf Acknowledgments}. }
We would like to thank Professor T.Hara of Kyushu university 
who proposed to us to publish this note. K.R.Ito would like to thank 
the Grand-in-Aid for Scientific Research (C) 15540222 from JSPS.


\end{document}